\documentclass[aps,amssymb,12pt,showpacs,showkeys,letterpaper]{revtex4-1}
\usepackage{graphicx}
\usepackage{amsmath}
\usepackage{epsfig}
\usepackage{bm}

\newcommand{\be}{\begin{equation}}
\newcommand{\ee}{\end{equation}}
\newcommand{\beq}{\begin{eqnarray}}
\newcommand{\eeq}{\end{eqnarray}}

\usepackage{amsmath,amssymb}

\begin{document}

\title{The Kepler problem in the Snyder space }

\author{ Carlos Leiva }
 \email{cleivas@uta.cl}
\affiliation{\it Departamento de F\'{\i}sica, Universidad de
Tarapac\'{a}, Casilla 7-D, Arica, Chile}

\author{ Joel Saavedra }
 \email{joel.saavedra@mail.ucv.cl}

\affiliation{\it Instituto de F\'{\i}sica, Pontificia Universidad
de Cat\'{o}lica de Valpara\'{\i}so,\\ Av. Universidad 330,
Curauma, Valpara\'{\i}so , Chile}

\author{ J. R. Villanueva }
 \email{jrvillanueval@uta.cl}
\affiliation{\it Departamento de F\'{\i}sica, Universidad de
Tarapac\'{a}, Casilla 7-D, Arica, Chile}

\date{\today}

\begin{abstract}

 In this paper we study the Kepler problem in the non
 commutative Snyder  scenario. We characterize the  deformations in the Poisson
 bracket algebra under a mimic procedure from quantum standard formulations and taking
 into account a general recipe to build
  the noncommutative phase space coordinates  (in the sense of  Poisson brackets).
We obtain an expression to
 the deformed potential, and then  the consequences in the precession
  of the orbit of  Mercury are calculated. This result allows us to
 find an estimated value for the non commutative deformation parameter
 introduced.

\end{abstract}

\pacs{04.20.-q, 04.70.Bw, 04.90.+e}

\keywords{}

\maketitle
\section{\label{sec:Int}Introduction}

Non commutativity has became  a serious fellow among the physics
theories, since minimal fundamental lengths  have been introduced
by the leading theories of loop quantum gravity and string theory.
This minimal fundamental length usually is identified as the
Planck length and it is supposed that under that scale Physics is
totally different, even from the standard Quantum Physics.

There are many ways to introduce non commutativity, usually the
Heisenberg algebra is deformed through a matrix that encodes the
lack of commutativity between the position operators. This way is
incompatible with Lorentz symmetry and many difficulties arise due
to the many changes that  abandoning this fundamental symmetry
implies. But there is a  safer way. In fact, H. Snyder in the
$40's$ \cite{Snyder} proposed a modification of the Heisenberg
algebra that implies  discrete spectra of the spacetime operators.
This modification is included among the $\kappa$-deformed spacetime
modifications.

In fact, the noncommutative spacetime program was forgotten due
the successfully renormalization program in the standard model.
However there is a renewed interest  due to  the develop of loop
quantum gravity and string theories with their discrete
spacetimes.

One of the problems with the leading theories of quantum gravity
today is the lack of experimental confirmations. In that
direction, this paper shows a possible way to measure the
implications of a non commutative spacetime, using the well known
Kepler celestial mechanics; introducing a deformation parameter in
the Kepler potential and forecasting deformations in the orbits of
planets. There are some previous efforts dealing with this
problem, but they used a non commutativity that is not compatible
with Lorenz symmetry, and that is a very undesirable
feature,\cite{MirKasimov:1996hv}\cite{mai}\cite{Mirza}.

The paper is organized as follows: in the next section, a short
review of non commutative algebras is given,  in the third section
the Kepler problem in the Snyder spacetime is developed obtaining
an advance of perihelion of a planet due to the deformed
considerations and, finally conclusions are given in the last
section.

\section{\label{sec:Dilatonic}Non commutative algebras}

\subsection{General case}

 In a $(n+1)$ dimensional Minkowski spacetime, we
introduce the non commutativity through:

\begin{equation}
[\bar{x}_\mu,\bar{x}_\nu]=lM_{\mu\nu},
\end{equation}

\noindent where $\bar{x}$ is the non commutative coordinate and $l$ a
parameter measuring the non commutativity with dimension of
squared length, usually identifying $\sqrt{l}$ with $l_p$, the
Planck longitude and $M_{\mu\nu}$ the rotations generator.

It is usual to demand that the Poincar\'{e} algebra is untouched,
then we have the standard commutations relations

\begin{eqnarray}
[M_{\mu \nu},M_{\rho \sigma}]= \eta_{\mu \nu}M_{\mu \nu}-\eta_{\mu
\rho}M_{\nu \sigma}-\eta_{\nu \sigma}M_{\mu \rho}+\eta_{\mu
\sigma}M_{\nu \rho},
\end{eqnarray}
\begin{equation}
 [p_{\mu} , p_{\nu}]=0. \nonumber
\end{equation}

 We can obtain a general expression for the new
coordinates taking \cite{Bat}

\begin{equation}
\bar{x}_\mu=x_\mu \phi_1(A)+l(xp)p_\mu \phi_2(A), \label{genreal}
\end{equation}

\noindent where $\phi_1$ and $\phi_2$ are two dependent functions of the
quantity $A=sp^2$, and the relation between them is

\begin{equation}
\phi_2=\frac{1+2\phi_1' \phi_1}{\phi_1-2A\phi_1'},
\end{equation}

\noindent where $(')$ denotes derivative respect to $A$.

 We have freedom to take any value of $\phi_1$ in order to
obtain the realization of the non commutativity, the only
restriction is the boundary condition $\phi(0)=1$,  to retrieve
the ordinary commutativity.

In the general case the commutator between coordinates and momenta
is

\begin{equation}
[\bar{x_\mu},p_\nu]=i(\eta_{\mu \nu}\phi_1+lp_\mu p_\nu \phi_2).
\end{equation}

\subsection{Snyder case}

Among the infinite possibilities  choosing the value of $\phi_1$,
there is a very special case: taking $\phi_1=1$. This choice
implies that $\phi_2=1$, that leads to the so called Snyder space,
characterized by

\begin{equation}
[x_\mu,x_\nu]=ilM_{\mu \nu}, \label{nc}
\end{equation}
\begin{equation}
[x_\mu ,p_\nu]=i\delta_{\mu \nu}-ilp_\mu p_\nu, \label{nl}
\end{equation}

\begin{equation}
[p_\mu,p_\nu]=0.
\end{equation}

This is a very interesting case and many works have investigated
about it since the Snyder's paper itself \cite{Snyder},  \cite{
MirKasimov:1996hv}, \cite{Chatterjee:2008bp} and others.

\section{\label{sec:Dilatonic}The kepler problem in the Snyder non commutative Euclidian
space}

Classical euclidian $n$ dimensional  Snyder Space is characterized
by its non linear commutation relations (in the sense of Poisson
brackets), between the variables of the phase space. They can be
set following the inverse of Dirac quantization recipe

\begin{eqnarray}
\{x_i,x_j\}&=& l_{p}^{2}L_{ij}, \\ \label{cnc}
\{x_i,p_j\}&=&\delta_{ij}-l_{p}^{2}p_ip_j,\\ \label{cnl} \{p_i,p_j\}&=&0,
\end{eqnarray}
where $l_p$ is the Planck longitude and measures the deformation
introduced in the canonical Poisson brackets, and $L_{ij}$ is
defined as a dimensionless matrix proportional to the angular
momentum.

 The Kepler potential $V=-\frac{\kappa}{\sqrt{x_i^2}}$ is implemented in the general non
 commutative case, taking

 \begin{equation}
 V(\bar{x})=-\frac{\kappa}{\sqrt{\bar{x}_i \bar{x}_i}},
 \end{equation}

\noindent and considering the recipe from (\ref{genreal}), we obtain at the first
order in $l$

\begin{equation}
V(x)=-\frac{\kappa}{\sqrt{x_i^2\phi_1^2+2l_{p}^{2}(xp)^2\phi_1\phi_2 }}.
\end{equation}

For the Snyder realization ($\phi_1=\phi_2=1$), we have that

\begin{equation}
V(x)=-\frac{\kappa}{\sqrt{x_i^2+2l_{p}^{2}(xp)^2}},
\end{equation}

\noindent so, using polar coordinates for a plane motion,

\begin{eqnarray}
x&=&\rho \hat{\rho}, \\
p&=&m(\dot{\rho}\hat{\rho}+\rho \dot{\theta}\hat{\theta}),
\end{eqnarray}

\noindent the Lagrangian for a particle in the Snyder-Kepler potential can
be written as

\begin{equation}
\mathcal{L}=\frac{1}{2}m\left[1-\frac{2l_{p}^{2}km}{\rho}\right]\dot{\rho}^2+\frac{1}{2}m\rho^2
\dot{\theta}^2+\frac{k}{\rho}.
\end{equation}

We still have the angular momentum $L=m\rho^2\dot{\theta}$ as a
constant of motion, so considering a particle with  energy $E$ we
obtain for the radial equation

\begin{equation}
\dot{\rho}^2=\frac{2}{mf(\rho)}\left[
E- V_{cl}(\rho) \right],
\end{equation}

\noindent where $f(\rho)=(1+\frac{2\kappa l_{p}^{2}m}{\rho})$ and $V_{cl}(\rho)=\frac{L^2}{2m\rho^2}-\frac{\kappa}{\rho}$, is the classical effective potential for the two-bodies problem. In this sense, our interest is to study the non-commutative correction to the confined orbit, so the constant of motion $E$ is restricted to the values

$$
0> E >E_c\equiv-\frac{ \kappa}{2 \rho_c},
$$

\noindent where  $\rho_c =(m \kappa)^{-1} L^2$ is the radius of
the circular orbit and $E_c$ is the energy at this point. Now, we
can write a dimensionless equation of motion in terms of these
quantities as

\begin{equation}
(-x')^2=\left(2x-x^2-\mathfrak{E}\right)\left(1+2 J^2 x\right)^{-1},
\label{era}
\end{equation}

\noindent where $J=(m \kappa l_p)/L$, $1>\mathfrak{E}\equiv E/E_c >0$,  $x_- \geq x \equiv \rho_c / \rho \geq x_+$ (with $x_{\pm} \equiv
\rho_c / \rho_{\pm} = 1 \mp \sqrt{1-\mathfrak{E}}$), and $x'=dx/d\theta$. Performing the substitution $x = A -y$, with $A=(4 J^2-1)(6 J^2)^{-1}$,  eq. (\ref{era}) becomes

\begin{equation}
y'^2=\frac{1}{8J^2}\frac{ 4 y^3 - g_2 y - g_3}{(h-y)^2},
\label{ef}
\end{equation}

\noindent where $h=(1+2J^2)(3J^2)^{-1}$, and the invariants are given by

$$
g_2 = \frac{1+4J^2+4J^4( 4-3\mathfrak{E})}{3 J^4},\quad \textrm{and}\quad
g_3=\frac{(1+2J^2)[1+4J^2-4J^4( 8-9\mathfrak{E})]}{27 J^6}.
$$

Therefore, choosing $\theta =0$ at $y = y_+$ and integrating eq.
(\ref{ef}),  we find

\begin{equation}
\frac{\theta}{\sqrt{8J^{2}}}= W(y_+)-W(y),
\label{solf}
\end{equation}
\noindent where
\begin{equation}
W(y)=(C - y)\wp(y; g_2,
g_3)-\zeta(y; g_2, g_3),
\end{equation}
\noindent  where $\wp$ is the Weierstrass-p function, and $\zeta$
is the Weierstrass-z function. Equation (\ref{solf}) represents
the formal solution for the Kepler's problem when the
non-commutativity is taken into account. But we still can say
something more about the deformation parameter, $l_p$. To do this,
we study the advance of perihelion starting from (\ref{era}),
expanding to order $J^2$, and neglecting $x^3$ terms. Thus, we
obtain

\begin{eqnarray}
\left(-\frac{d x}{d\theta}\right)^2 &&\approx -\mathfrak{E} + 2 (1+J^2 \mathfrak{E})x-(1+4J^2)x^2 \nonumber \\
&& = \frac{(1+J^2 \mathfrak{E})^2}{(1+4J^2)}-\mathfrak{E}-(1+4J^2)\left(x-\frac{(1+J^2 \mathfrak{E})}{(1+4J^2)}\right)^2,
\label{era2}
\end{eqnarray}

\noindent so, it yields

\begin{equation}
x\equiv \frac{\rho_c}{\rho} = C_1 + C_2 \cos (k \theta + \theta_0),
\end{equation}

\noindent where

$$
C_1=\frac{1+J^2 \mathfrak{E}}{1+4J^2},\quad
C_2=k^{-1} \left(\frac{(1+J^2 \mathfrak{E})^2}{1+4J^2}-\mathfrak{E}\right)^{1/2}, \quad
k=\sqrt{1+ 4 J^2}.
$$

Therefore, the correction for the  advance of perihelion is given
by

\begin{equation}
\Delta \theta = \frac{2\pi}{k}=2\pi\left(1+4J^2\right)^{-1/2},
\end{equation}

\noindent which can be approximated as a deviation of the
Newtonian orbit

\begin{equation}
\Delta \theta \simeq 2\pi\left(1- 2J^2\right)= 2\pi+\delta\theta_{nc},
\end{equation}

\noindent where $\delta\theta_{nc}=-4\pi J^2$ is the
non-commutative correction. In order to obtain the value of the
deformation parameter, we can consider that the discrepancy of the
observational data and the theoretical value in the specific case
of Mercury (see TABLE \ref{t1}), could be  due to the
non-commutativity scenario. Obviously we choose Mercury because it
is an usual natural laboratory to check deformations. This is
because it is expected that any little effect can be observable in
its orbit due to Mercury is the nearest planet to the sun.
Therefore, we obtain $l_p =1.68\times 10^{-32}$.

\begin{table}[ht]
\caption{Sources of the precession of perihelion for Mercury}
\begin{tabular}{|c|c|}
  \hline
  \textbf{Amount (arcsec/Julian century)} & \textbf{Cause} \\ \hline
  $5028.83\pm 0.04$\,\cite{nasa} & \textrm{Coordinate (due to the precession of the equinoxes)} \\ \hline
  $530$\, \cite{matzner} & \textrm{Gravitational tugs of the other planets} \\ \hline
  $0.0254$ & \textrm{Oblateness of the Sun (quadrupole moment)} \\ \hline
  $42.98\pm 0.04$\, \cite{Iorio:2004ee} & \textrm{General Relativity} \\ \hline
  $5603.24$ & \textrm{Total} \\ \hline
  $5599.7$ & \textrm{Observed} \\ \hline
  $-3.54$ & \textrm{Discrepancy} \\
  \hline
\end{tabular}
\label{t1}
\end{table}
\section{Final Remarks}
In this article we have described the effects of Snyder space non
commutativity on  Kepler problem and have studied its effect on a
planetary orbit. We introduced non commutativity  performing the
deformations in Poisson bracket algebra under a mimic procedure
from quantum standard formulations and then, using  a general
recipe to build  the noncommutative phase space coordinates
 (in the sense of  Poisson brackets). We found that the deformation in
 the central potential  allows us to write a Lagrangian for a particle in the Snyder-Kepler
 potential and to obtain the formal solution for the Kepler's problem when the non-commutativity is
taken into account. Our solution is given in terms of
Weierstrass-p ($\wp$)  and Weierstrass-z ($\zeta$) functions.
Then, used our analytical results to compute the advance of
perihelion of an planetary orbit. In fact, we found that is given
by $\Delta \theta =
\frac{2\pi}{k}=2\pi\left(1+4J^2\right)^{-1/2}$, which can be
approximated as a deviation of the Newtonian orbit as follow $
\Delta \theta \simeq 2\pi\left(1- 2J^2\right)=
2\pi+\delta\theta_{nc}$ where $\delta\theta_{nc}=-4\pi J^2$ is the
non-commutative correction.

Finally  we applied this formula to fix  the discrepancy between
observational data and the theoretical value obtined from
different classical sources and, under the hypothesis that the
discrepancy is due to ntking into account the non commuttivity of
the space, we obtained an estimated value for the non commutative
deformation parameter given by $l_p =1.68\times 10^{-32}$. In the
future we would like to see the value of deformation parameter in
more general setting as the advance of perihelion in the neighbor
of black hole, we left this issue for a future work.
\begin{acknowledgments}
C. L. was supported by Universidad de Tarapac\'{a} Grant 4723-11.
J. V was supported by Universidad de Tarapac\'{a} Grant 4720-11.
J.S. was supported by COMISION NACIONAL DE CIENCIAS Y TECNOLOGIA
through FONDECYT Grant 1110076, 1090613 and 1110230. This work was
also partially supported by PUCV DI-123.713/2011.
\end{acknowledgments}

\end{document}